\documentclass[11pt]{article}
\usepackage{graphicx}
\usepackage{epstopdf}
\usepackage{amsmath, amssymb}
\usepackage{shuffle}

\usepackage{hyperref}
\hypersetup{bookmarksopen,bookmarksnumbered,pdfstartview=FitH,colorlinks,citecolor=blue,unicode}
\usepackage[numbers,sort&compress]{natbib}

\def\eq{\begin{equation}}
\def\eqe{\end{equation}}
\def\eqa{\begin{eqnarray}}
\def\eqae{\end{eqnarray}}
\def\G{\Gamma}

\def\f#1#2{{\textstyle{#1\over#2}}}

\begin{document}

\title{{\bf\color{blue} Factorization of Chiral String Amplitudes}\\[-1.2in]
{\normalsize March 8, 2016\hfill YITP-SB-16-6}\\[1in]}
\date{}

\author{Yu-tin Huang$^{a,b}$\footnote{\href{yutin@phys.ntu.edu.tw}{yutin@phys.ntu.edu.tw}}\, , Warren Siegel$^{c}$\footnote{\href{mailto:siegel@insti.physics.sunysb.edu}{siegel@insti.physics.sunysb.edu}\vskip 0pt \hskip 8pt \href{http://insti.physics.sunysb.edu/\~siegel/plan.html}{http://insti.physics.sunysb.edu/$\sim$siegel/plan.html}}\, , Ellis Ye Yuan$^{b}$\footnote{\href{mailto:yyuan@ias.edu}{yyuan@ias.edu}}\bigskip\\ $^a$  \emph{Department of Physics, National Taiwan University, Taipei 10617, Taiwan}\\
$^b$\emph{School of Natural Sciences, Institute for Advanced Study, Princeton, NJ 08540, USA}
\\$^c$ \emph{C.~N.~Yang Institute for Theoretical Physics}\\ \emph{State University of New York, Stony Brook, NY 11794-3840}}
\maketitle

\begin{abstract}

\normalsize

We re-examine a closed-string model defined by altering the boundary conditions for one handedness of two-dimensional propagators in otherwise-standard string theory.  We evaluate the amplitudes using Kawai--Lewellen--Tye factorization into open-string amplitudes.  The only modification to standard string theory is effectively that the spacetime Minkowski metric changes overall sign in one open-string factor.  This cancels all but a finite number of states:  As found in earlier approaches, with enough supersymmetry (e.g., type II) the tree amplitudes reproduce those of the massless truncation of ordinary string theory.  However, we now find for the other cases that additional fields, formerly thought to be auxiliary, describe new spin-2 states at the two adjacent mass levels (tachyonic and tardyonic).
The tachyon is always a ghost, but can be avoided in the heterotic case.

\end{abstract}

\newpage

\section{Introduction}

Some time ago a new method was proposed for expressing complete particle scattering amplitudes in arbitrary dimensions using an integral expression similar to those found in string theory \cite{Cachazo:2013hca,Cachazo:2013iea,Cachazo:2014nsa,Cachazo:2014xea}, but with the insertion of delta-functions enforcing the ``scattering equation" constraints obtained semiclassically from such string amplitudes \cite{Gross:1987kza}.  String-like theories were proposed that reproduced these amplitudes, at least in the cases where Yang--Mills factors (in spectrum/vertices) were supersymmetric \cite{Mason:2013sva,Ohmori:2015sha,Casali:2015vta}.  There the delta-functions arose from related delta-functions inserted explicitly into the vertex operators. Interesting relations between such representations of field theory amplitudes and amplitudes in ordinary string theories were also explored in \cite{Bjerrum-Bohr:2014qwa}.

Recently a new derivation was given from standard string theories by two simple modifications~\cite{Siegel:2015axg}:  (1) A singular worldsheet gauge was used for the string Lagrangian to (almost) trivialize the dependence on the worldsheet coordinate $\bar z$; in this singular limit the string had only one handedness.  Earlier this limiting model was used to make T-duality manifest, and calculate $\alpha'$ corrections to the effective action in a way that preserved it~\cite{Hohm:2013jaa}.  (2) The boundary conditions on worldsheet propagators (especially for the bosons) were changed in a way consistent with this interpretation (essentially a Bogoliubov transformation), and with a corresponding modification of the Virasoro constraints such that the spectrum was apparently truncated to the massless sector.  The result of these two modifications was that integration over the trivial $\bar z$ dependence automatically produced the scattering-equation delta-functions, while standard string vertex operators were used.  But the method was successful only with the same restrictions as in the earlier string-like theories.

In this paper we give an alternative approach to this string theory:  The modified boundary conditions are still used to define the model from standard string theories.  However, since final results are gauge independent, we evaluate amplitudes in the usual conformal gauge.  Then the scattering-equation delta-functions are not seen explicitly.  As an alternative method of evaluating the integrals, we use instead the factorization of closed-string amplitudes into open-string ones~\cite{Kawai:1985xq,BjerrumBohr:2010hn}, which also treats $z$ and $\bar z$ integration as independent.  (Essentially one Wick rotates back to a worldsheet with Minkowski metric.)  The change in the sign of the $\bar z$ part of the $X$ propagator, implied by the change in boundary conditions, means that the main change in this evaluation of the amplitude is that the spacetime metric undergoes an effective overall sign change in one of the open-string factors.  One result is that the singularities almost completely cancel, leaving rational functions of momenta.  

This reproduces the massless amplitudes found previously, but it also consistently evaluates the amplitudes for less supersymmetry:  Additional fields, previously thought to be auxiliary\ \cite{Hohm:2013jaa}, are now found to describe a pair of new spin-2 states at positive and/or negative $m^2$, and their spin-3/2 superpartners when available.  (They return to being auxiliary in the limit $m^2\to\infty$.)  The tachyons are also ghosts.  The bosonic string has both signs of masses, type II has neither, and heterotic has a choice of just one of physical tardyons or ghost tachyons.

\section{Amplitudes with Chiral Boundary Conditions}\label{sec2}

\subsection{Review}\label{sec21}

In \cite{Hohm:2013jaa} a modification of string theory was suggested that made T-duality manifest, and apparently restricted it to massless states.  (We won't consider T-duality explicitly here, as it is spontaneously broken to Lorentz symmetry for the purposes of perturbation theory.)  It was used to construct a T-duality invariant effective action for the massless fields, including complete $\alpha'$ corrections, by a method equivalent to $\beta$-function techniques.  The fields included the usual states from left$\otimes$right states of open strings, but T-duality required also left$\otimes$left and right$\otimes$right.  However, the fields for the latter states were treated as auxiliary in the action, by making an $\alpha'$ expansion.  (Scattering amplitudes weren't considered there.  Only the bosonic string was treated; generalization to supersymmetric strings was obvious, but expected to trivialize $\alpha'$ corrections.)

In \cite{Siegel:2015axg} this model was generalized to amplitudes by explaining the apparent restriction to massless states through the change in the boundary conditions.
(The trivialization of $\bar z$ dependence came from a singular choice of gauge, which we will not need in this paper.)  In particular, the change of boundary condition for $X$ is implemented by adding a homogenous solution to the propagator,
\eq
\langle XX\rangle \to  -\ln(\bar zz) +\ln(\bar z^2) = \ln\left(\frac{\bar{z}}{z}\right)
\eqe
(where the first term has the usual $i\epsilon$ prescription).  Effectively this changes the overall sign of the spacetime Minkowski metric associated with the $\bar z$ sector.  (By BRST and worldsheet supersymmetry, it is seen to change the sign of fermionic and ghost propagators in the $\bar z$ sector as well.)

In oscillator language, this change in the boundary conditions corresponds to a Bogoliubov transformation that switches creation and annihilation operators for $\bar z$ modes,
\eq
\bar a \to \bar a^\dag , \quad \bar a^\dag \to -\bar a
\eqe
The effect on the Virasoro operators is then
\eq
L_0 = \alpha'(\f12 p)^2 +N-1, \quad \bar L_0 \to -\alpha'(\f12 p)^2 +\overline N-1
\eqe
($N$ and $\overline N$ are the level-number operators, taking values 0,1,...), and thus the constraints
\eq
L_0 +\bar L_0 \to N+\overline N-2 = 0, \quad L_0 -\bar L_0 \to \f12\alpha' p^2 +N-\overline N = 0, \quad 
\eqe
restrict the spectrum to levels and masses
\eq
(N,\overline N;\f14\alpha' m^2) = (1,1;0), (2,0;1), (0,2;-1)
\eqe
However, based on \cite{Hohm:2013jaa}, these massive states were assumed to still be auxiliary.
As we'll see later, that treatment is no longer possible when four-and-higher-point amplitudes are considered (except perhaps in the limit where those masses become infinite).

Such states do not appear if both left and right factors have massless ground states, from supersymmetry or internal symmetry, as from type II superstrings or single-trace heterotic super Yang--Mills amplitudes.  So more generally we can write
\eq
L_0 = \alpha'(\f12 p)^2 +N_L, \quad \bar L_0 = -\alpha'(\f12 p)^2 +N_R \quad\Rightarrow\quad N_L + N_R = 0 , \quad \f14\alpha' m^2 = \f12(N_L - N_R)
\eqe
but where each of $N_{L,R}$ starts at $-1$ if bosonic or at 0 if supersymmetric (or internal symmetry scalars of heterotic).
For example, for the supergravity sector of the heterotic string, there is only one massive spin 2, but either tachyonic or tardyonic depending on which sector is supersymmetric.  (Both sectors have such an $N=+1$, but only one has the corresponding $-1$.)  In the heterotic Yang--Mills case, the restriction to single traces for group theory eliminates the nonplanar group theory factors corresponding to supergravity intermediate states.

The equivalent analysis in conformal field theory language is that worldsheet derivatives still contribute in the usual way to the conformal weight of vertex operators, but the ``$k^2$" contribution comes with opposite sign for $\bar L_0$.  So vertex operators are constructed in the usual string way, but the choice of mass and left and right level numbers is modified as described above.

The net effect in amplitudes is that the Koba--Nielsen factor becomes
\eq
|z_{ij}|^{4\,k_i\cdot k_j}\, |\bar{z}_{ij}|^{4\,k_i\cdot k_j}\rightarrow |z_{ij}|^{4\,k_i\cdot k_j}\, |\bar{z}_{ij}|^{-4\,k_i\cdot k_j}.
\eqe
To simplify notation we set $\alpha'=4$ above and in the rest of this paper. Similarly, any inner product of polarization vectors with themselves or momenta that results from the contraction of right-movers will also obtain a minus sign.

\subsection{Effect on the Kawai--Lewellen--Tye Relation}

In this paper we investigate closed string amplitudes using their factorized representation, i.e., the Kawai--Lewellen--Tye (KLT) relation. In this setup the amplitudes in the chiral string is identical to a simple operation acting on those in the standard closed string.

We reserve the notation $M$ for closed string amplitudes and $A$ for open strings. Schematically a standard closed string amplitude can be expressed as
\begin{equation}
M=\mathbf{A}\mathbf{S}\tilde{\mathbf{A}}.
\end{equation}
Here $\mathbf{A}$ denotes a vector of open string amplitudes of size $(n-3)!$
\begin{equation}
\mathbf{A}=(A[\sigma_1],A[\sigma_2],\ldots,A[\sigma_{(n-3)!}])^{\mathrm{T}},
\end{equation}
whose elements are labeled by the ordering of external points on the disk boundary $\sigma_i\in S_n$. The KLT momentum kernel, $\mathbf{S}$, is a matrix depending only on the Mandelstam invariants. The choice of $\mathbf{A}$ (and $\tilde{\mathbf{A}}$, respectively) is not unique, and the form of $\mathbf{S}$ depends on the specific choice.

The two sets of open string amplitudes $\mathbf{A}$ and $\tilde{\mathbf{A}}$ are associated with the left- and right-moving sectors of the closed string, and so the chiral boundary condition affects only $\tilde{\mathbf{A}}$.

To illustrate this, consider a toy integral at four points
\begin{equation}\label{toyintegral}
I:=\int\mathrm{d}^2z\,\frac{|z|^{-2s}|1-z|^{-2t}}{z(1-z)\bar{z}}
=\int d^2z\,\frac{z^{-s}(1-z)^{-t}}{z(1-z)}\,\frac{\bar{z}^{-s}(1-\bar{z})^{-t}}{\bar{z}},
\end{equation}
which mimics the scattering of four massless scalars with stringy corrections, and $s,t,u$ are the standard Mandelstam inariants at four points. To translate this into a KLT representation, recall that $z=\tau+i\sigma$:
We Wick rotate from complex Euclidean coordinates to real Minkowskian (lightcone) coordinates $z\mapsto\xi$ and $\bar{z}\mapsto\eta$ in \eqref{toyintegral}, and integrate $\xi$ and $\eta$ along their own real axes (with appropriate $i\epsilon$ prescriptions). By folding the $\eta$ contour at the branch points of the integrand, one completely separates the left- and right-movers into two individual real integrals
\begin{equation}
\begin{split}
I&=\left(\int_0^1\mathrm{d}\xi\,\xi^{-s-1}(1-\xi)^{-t-1}\right)\sin(\pi t)\left(\int_1^{\infty}\mathrm{d}\eta\,\eta^{-s-1}(\eta-1)^{-t}\right)\\
&=\frac{\Gamma(-s)\Gamma(-t)}{\Gamma(u)}\,\sin(\pi t)\,\frac{\Gamma(-u)\Gamma(1-t)}{\Gamma(1+s)}.
\end{split}
\end{equation}
In this case $\mathbf{S}=\sin(\pi t)$, which the monodromy resulting from folding the $\eta$ contour.

Note that the monodromy in $\mathbf{S}$ is not affected after imposing the chiral boundary condition, because it is only related to the sign change of the combination $\xi\eta$ and $(1-\xi)(1-\eta)$ along the original $\xi$- and $\eta$-contour, and the chiral boundary condition merely switches
$\xi\eta\to\xi/\eta$ and $(1-\xi)(1-\eta)\to(1-\xi)/(1-\eta)$.
Consequently it only modifies the final $\eta$ integral by flipping the sign of the flat spacetime metric therein, $\eta_{\mu\nu}\to-\eta_{\mu\nu}$. It is a simple calculation to check that
\begin{equation}
I^{\rm chiral}=\frac{\Gamma(-s)\Gamma(-t)}{\Gamma(u)}\,\sin(\pi t)\,\left[\frac{\Gamma(-u)\Gamma(1-t)}{\Gamma(1+s)}\right]_{\rm flip}=\frac{\pi}{s}.
\end{equation}
using $\sin(\pi t)=\pi/\Gamma(t)\Gamma(1-t)$.  We see that this flipping operation projects away all the massive poles in the original integral, leaving just a rational function. More amusingly, this result is exactly the leading term of the original integral $I$ in its $\alpha'$ expansion!

This is the first demonstration of how the truncation of the closed string spectrum following from the Virasoro constraints with the new boundary condition holds in amplitudes at \textit{generic} $\alpha'$. 
The above flipping operation directly applies to more general cases, i.e., we always flip the sign of the spacetime metric $\eta_{\mu\nu}$ in $\tilde{\mathbf{A}}$
\begin{equation}
M^{\rm chiral}=\mathbf{A}(\eta_{\mu\nu})\mathbf{S}[\tilde{\mathbf{A}}(\eta_{\mu\nu})]_{\rm flip}=\mathbf{A}(\eta_{\mu\nu})\mathbf{S}\tilde{\mathbf{A}}(-\eta_{\mu\nu}).
\end{equation}
From here on, we will refer to amplitudes evaluated with the new chiral boundary conditions as chiral amplitudes, denoted in the superscript.

\section{Amplitudes in the Superstring}

\subsection{The general result}

As discussed in Section \ref{sec2}, in type II superstrings the chiral boundary condition truncates the string spectrum to only the massless states at generic $\alpha'$, and so we expect that the closed string amplitudes computed therein are exactly the supergravity amplitudes
\begin{equation}\label{superM}
M_{\rm super}=\mathbf{A}_{\rm super}^{\mathrm{T}}\mathbf{S}\tilde{\mathbf{A}}_{\rm super}\quad\longrightarrow\quad
M_{\rm super}^{\rm chiral}=M_{\rm sugra}=\mathbf{A}_{\rm super}^{\mathrm{T}}\mathbf{S}[\tilde{\mathbf{A}}_{\rm super}]_{\rm flip}.
\end{equation}

This statement can be reformulated into a simpler form. The open superstring amplitudes admit an expansion onto a basis of super Yang--Mills (SYM) amplitudes \cite{Mafra:2011nv,Mafra:2011nw}
\begin{equation}
\mathbf{A}_{\rm super}=\mathbf{F}\mathbf{A}_{\rm SYM},
\end{equation}
where the vector of SYM amplitudes $\mathbf{A}_{\rm SYM}$ is again of size $(n-3)!$, and $\mathbf{F}$ is a matrix that absorbs all the disk integrals and depends only on the Mandelstam invariants. Let us denote $\mathbf{S}_0$ as the leading part of the KLT kernel $\mathbf{S}$ in its $\alpha'$ expansion. Then the KLT relation is
\begin{equation}
M_{\rm sugra}=\mathbf{A}_{\rm SYM}^{\mathrm{T}}\mathbf{S}_0\tilde{\mathbf{A}}_{\rm SYM}.
\end{equation}
Note that the SYM amplitudes have uniform total degree in terms of the Lorentz-invariant products, and so flipping $\tilde{\mathbf{A}}_{\rm SYM}$ only results in a possible overall sign, which is not important. Hence the statement \eqref{superM} is equivalent to
\begin{equation}\label{susystatement}
\mathbf{F}^{\mathrm{T}}\mathbf{S}[\tilde{\mathbf{F}}]_{\rm flip}=\mathbf{S}_0.
\end{equation}

From now on we choose to identify the basis used on the left with that on the right
\begin{align}
\mathbf{A}_{\rm super}=\tilde{\mathbf{A}}_{\rm super}&=(A_{\rm super}[1,\sigma,n-1,n])^{\mathrm{T}},\qquad \sigma\in S_{n-3},\\
\mathbf{A}_{\rm SYM}=\tilde{\mathbf{A}}_{\rm SYM}&=(A_{\rm SYM}[1,\sigma,n,n-1])^{\mathrm{T}},\qquad \sigma\in S_{n-3}.
\end{align}
With this convention we explicitly verified \eqref{susystatement} at four and five points. The detailed expressions for $\mathbf{S}$ and $\mathbf{F}$ (in terms of $\Gamma$ and $_3F_2$) used in theses checks can be found in~\cite{Schlotterer:2012ny}.

\subsection{Derivation}
In this subsection we provide a derivation for the statement \eqref{susystatement}. We start by reviewing some useful properties of the matrices $\mathbf{F}$ and $\mathbf{S}$.

Despite the complexity of the elements in the matrix $\mathbf{F}$ in general, it can be analyzed in terms of an expansion onto (a chosen {\bf Q} basis of) multi-zeta values (MZVs). In this context it was observed in \cite{Schlotterer:2012ny} that $\mathbf{F}$ admits a factorized form
\begin{equation}
\mathbf{F}=\mathbf{P}\mathbf{M},
\end{equation}
where
\begin{equation}
\mathbf{P}:=\sum_{k=0}^{\infty}f_2^k\mathbf{P}_{2k},\qquad
\mathbf{M}:=\sum_{p=0}^{\infty}\sum_{\substack{i_1,\ldots,i_p\\\in 2{\bf N}^++1}}f_{i_1}f_{i_2}\cdots f_{i_p}\mathbf{M}_{i_p}\cdots\mathbf{M}_{i_2}\mathbf{M}_{i_1},
\end{equation}
and $\mathbf{P}_{2k}$ and $\mathbf{M}_{2k+1}$ are the coefficients of the Riemann zeta values in $\mathbf{F}$
\begin{equation}
\mathbf{P}_{2k}:=\mathbf{F}\big|_{(\zeta_2)^k},\qquad
\mathbf{M}_{2k+1}:=\mathbf{F}\big|_{\zeta_{2k+1}},
\end{equation}
which are functions of the Mandelstam invariants\footnote{Examples of the $\mathbf{P}_{2k}$ and $\mathbf{M}_{2k+1}$ matrices can be found in \cite{BroedelWebsite}, and systematic methods in obtaining them were introduced in \cite{Broedel:2013tta,Broedel:2013aza}.}. The non-commutative word $f_{i_1}\cdots f_{i_p}$ formed by the $f$ symbols is a convenient bookkeeping for the MZVs to be associated to the corresponding matrix profuct \cite{Brown:2011ik}, and the summation in $\mathbf{M}$ is over all inequivalent words.

Two conjectures are proposed in~\cite{Schlotterer:2012ny}, regarding properties of the matrices $\mathbf{P}$ and $\mathbf{M}$ respectively. The first one states that $\mathbf{P}$ ``projects'' $\mathbf{S}$ into $\mathbf{S}_0$
\begin{equation}
\mathbf{P}^{\mathrm{T}}\mathbf{S}\mathbf{P}=\mathbf{S}_0.
\end{equation}
And the second one states that every $\mathbf{M}_{2k+1}$ ``commutes'' with $\mathbf{S}_0$
\begin{equation}
\mathbf{M}_{2k+1}^{\mathrm{T}}\mathbf{S}_0=\mathbf{S}_0\mathbf{M}_{2k+1},\qquad\forall k\in{\bf N}^+.
\end{equation}
These two conjectures were explicitly verified up to the order $\alpha'^{21}$ at five points, $\alpha'^{9}$ at six points, and $\alpha'^{7}$ at seven points \cite{Schlotterer:2012ny,Broedel:2013tta}.

Given this structure of the open superstring amplitudes, the statement \eqref{susystatement} reformulates to
\begin{equation}
[\mathbf{M}^{\mathrm{T}}\mathbf{P}^{\mathrm{T}}]_{\rm flip}\mathbf{S}\mathbf{P}\mathbf{M}=\mathbf{S}_0.
\end{equation}
Each $\mathbf{P}_{2k}$ has even degree in terms of the Mandelstam variables, and so $\mathbf{P}=\mathbf{P}_{\rm flip}$. By applying the two conjectures above we obtain an equivalent statement
\begin{equation}
\mathbf{M}_{\rm flip}\mathbf{M}=\mathbf{1}.
\end{equation}
Here note that when writing $\mathbf{M}_{\rm flip}$ we implicitly mean that the ordering of the multiplication of the $\mathbf{M}_{2k+1}$'s is also reversed as compared to that in $\mathbf{M}$, apart from the fact that it reverses the overall sign of each $\mathbf{M}_{2k+1}$.

The above product between the two $\mathbf{M}$ matrices needs to be understood with a grain of salt, due to the non-trivial map between the $f$ words and the MZVs. While the products between $\mathbf{M}_{2k+1}$'s are normal matrix products, the ``product'' between the $f$ words is defined as a shuffle
\begin{equation}
f_{i_1}\cdots f_{i_r}\shuffle f_{i_r+1}\cdots f_{i_{r+s}}=\sum_{\sigma\in S_{r,s}}f_{i_{\sigma(1)}}\cdots f_{i_{\sigma(r+s)}},
\end{equation}
where
\begin{equation}
S_{r,s}:=\{\sigma\in S_{r+s}|\sigma^{-1}(1)<\cdots<\sigma^{-1}(r)\text{ and }\sigma^{-1}(r+1)<\cdots\sigma^{-1}(r+s)\}.
\end{equation}

Now we can expand $\mathbf{M}_{\rm flip}\mathbf{M}$ in terms of products of $\mathbf{M}_{2k+1}$'s
\begin{equation}
\mathbf{M}_{\rm flip}\mathbf{M}=\mathbf{1}+\sum_{p=1}^{\infty}\sum_{\substack{i_1,\ldots,i_p\\\in 2{\bf N}^++1}} g_{i_1,i_2,\ldots,i_p}\mathbf{M}_{i_p}\cdots\mathbf{M}_{i_2}\mathbf{M}_{i_1}.
\end{equation}
Due to the flipping, the coefficients in front of each product is
\begin{equation}
g_{i_1,i_2,\ldots,i_p}=\sum_{r=0}^{p}(-1)^{p-r}f_{i_p}\cdots f_{i_{r+2}}f_{i_{r+1}}\shuffle f_{i_1}f_{i_2}\cdots f_{i_r}.
\end{equation}
And so the problem reduces to showing that
\begin{equation}
g_{i_1,i_2,\ldots,i_p}\equiv0,
\end{equation}
for any $p>0$ and any choice of $\{i_1,i_2,\ldots,i_p\}$ (they may not be distinct).

This can be proven by induction. First of all obviously we have
\begin{equation}
g_{i_1}=-f_{i_1}+f_{i_1}=0,\qquad\forall i.
\end{equation}
Now suppose that this identity holds up to level $p-1$, then (note that there are only two terms in which $f_{i_p}$ appears at the end, which obviously cancel)
\begin{equation}
\begin{split}
g_{i_1,i_2,\ldots,i_p}
&=\sum_{r=0}^{p-1}(-1)^{p-r}\sum_{s=0}^rf_{i_1}\cdots f_{i_s}f_{i_p}(f_{i_{p-1}}\cdots f_{i_{r+1}}\shuffle f_{i_{s+1}}\cdots f_{i_r})\\
&=-\sum_{s=0}^{p-1}f_{i_1}\cdots f_{i_s}f_{i_p}(\sum_{r=s}^{p-1}(-1)^{p-r-1}f_{i_{p-1}}\cdots f_{i_{r+1}}\shuffle f_{i_{s+1}}\cdots f_{i_r})\\
&=-\sum_{s=0}^{p-1}f_{i_1}\cdots f_{i_s}f_{i_p}g_{i_{s+1},\ldots,i_{p-1}}=0.
\end{split}
\end{equation}
This provides a non-trivial consistency check that the amplitudes in the type II case are the supergravity amplitudes.

\section{Amplitudes in the Bosonic and Heterotic Strings}

One can straight forwardly apply the analysis in the previous section to that of bosonic and heterotic closed strings
\begin{equation}
M_{\rm bos}=\mathbf{A}_{\rm bos}^{\mathrm{T}}\mathbf{S}\mathbf{A}_{\rm bos},\qquad
M_{\rm het}=\mathbf{A}_{\rm super}^{\mathrm{T}}\mathbf{S}\mathbf{A}_{\rm bos}.
\end{equation}
In~\cite{Huang:2016tag} it was conjectured, and explicitly verified up to seven-points, that the massless bosonic open string amplitudes $\mathbf{A}_{\rm bos}$ admits a similar expansion as that for $\mathbf{A}_{\rm super}$
\begin{equation}
\mathbf{A}_{\rm bos}=\mathbf{F}\mathbf{B}.
\end{equation}
Each element in the matrix $\mathbf{B}$ has the same algebraic properties as the tree-level Yang--Mills amplitudes, and reduces to $\mathbf{A}_{\rm YM}$ upon taking $\alpha'\rightarrow0$. Crucial for our discussion is the fact that $\mathbf{B}$ is a rational function that contains only massless and tachyon poles $(s+1)^{-1}$. Note that the product $\mathbf{S}\mathbf{F}$ produces inverse tachyon poles such that for the heterotic string the tachyon poles cancel, while for the bosonic closed string double tachyon poles are avoided.

Now we apply the flipping operation. Since \eqref{susystatement} still holds, the resulting amplitude becomes
\begin{eqnarray}
M_{\rm bos}^{\rm chiral}:\;&&\mathbf{B}\mathbf{S}_0[\mathbf{B}]_{\rm flip}\nonumber\\
M_{\rm het}^{\rm chiral}:\;&&\mathbf{A}_{\rm super}^{\mathrm{T}}\mathbf{S}_0[\mathbf{B}]_{\rm flip},\quad  (\mathbf{B})^{\mathrm{T}}\mathbf{S}_0[\mathbf{A}_{\rm super}]_{\rm flip}
\end{eqnarray}
For the bosonic string, we see that due to the flip the tachyon pole in $[\mathbf{B}]_{\rm flip}$ turns into that of the first massive excitation, and the amplitude now contains poles at $m^2=(\pm1,0)$, as anticipated from the constraint analysis in Section \ref{sec2}. For the heterotic string, besides the usual massless excitation, one has either the tachyon or the tardyon depending on whether it is the bosonic or the superstring in the flipped sector. Note that since taking $\alpha'\rightarrow0$, $\mathbf{B}$ reduces to $\mathbf{A}_{\rm YM}$, the leading contribution in the $\alpha'$ expansion is again the graviton tree amplitude.

As previously discussed, the massive bosonic states in the theory with chiral boundary conditions are spin-2 states. We will see this directly by inspecting the three-point and four-point interactions. 

\subsection{Three-point amplitudes}
As has long been known (and reproduced in the KLT representation), closed-string three-point amplitudes are tensor products of open-string vertices.  This result was extended to the bosonic chiral string in~\cite{Naseer:2016izx} by expansion of the effective action of\ \cite{Hohm:2013jaa}, after the appearance of Lorentz Chern-Simons terms was recognized~\cite{Hohm:2014eba,Hohm:2015mka}.  Here we obtain those results more directly, and extend them to the other chiral strings, by using the chiral extension of the direct-product analysis.

As a shorthand notation, we write the vertex factors as the terms in the action from which they follow:  For the open string
\eq
F^2 \equiv \epsilon_{1}\cdot\epsilon_2\,\epsilon_3\cdot k_{12}{+}\epsilon_{2}\cdot\epsilon_3\,\epsilon_1\cdot k_{23}{+}\epsilon_{3}\cdot\epsilon_1\,\epsilon_2\cdot k_{31} , \quad 
F^3 \equiv \epsilon_1\cdot k_{23}\,\epsilon_2\cdot k_{31}\,\epsilon_3\cdot k_{12}
\eqe
(where $k_{ij}\equiv k_i-k_j$), and for the closed string
\eq
R \equiv F^2 \times F^2 , \quad R^2 \equiv F^3 \times F^2 + F^2 \times F^3  , \quad 
CS \equiv F^3 \times F^2 - F^2 \times F^3  , \quad R^3 \equiv F^3 \times F^3
\eqe
where ``$R^2$" is the Gauss-Bonnet combination and ``$CS$" is the term coming from Lorentz Chern-Simons.

Then normal strings have the three-point amplitudes
\begin{eqnarray}
\hbox{bosonic:} & (F^2+F^3)\times(F^2+F^3) & = R + R^2 + R^3 \nonumber\\
\hbox{heterotic:} & (F^2+F^3)\times F^2 \ \hbox{or}\  F^2\times(F^2+F^3) & = R +\f12 R^2 \pm\f12 CS \nonumber\\
\hbox{type II:} & F^2\times F^2 & = R  \\
\end{eqnarray}
while chiral strings have (for massless states)
\begin{eqnarray}
\hbox{bosonic:} & (F^2+F^3)\times(F^2-F^3) & = R + CS - R^3 \nonumber\\
\hbox{heterotic:} & (F^2+F^3)\times F^2 \ \hbox{or}\  F^2\times(F^2-F^3) & = R \pm\f12 R^2 +\f12 CS \nonumber\\
\hbox{type II:} & F^2\times F^2 & = R \\
\end{eqnarray}
The minus signs in the second factor come from our sign flipping prescription.

\subsection{Four-point amplitudes}
Here we consider the four-point amplitude, which is useful in determining the nature of the states in the spectrum. For later convenience, we give the massless open bosonic-string amplitude as:
\eq
A_{\rm 4,bos}(s,t)=\frac{\G(-s)\G(-t)}{\G(1+u)}(K^{ss}+K)
\eqe
The functions $K^{ss}$ and $K$ are gauge invariant rational functions of polarization vector and momentum inner products. Each is permutation invariant and schematically given by\footnote{Their explicit form can be found in~\cite{Kawai:1985xq}. Note that here they are defined with an explicit $(1-\alpha' s)(1-\alpha' t)(1-\alpha' u)$ pulled out, and an extra minus sign in front of $K^{ss}$ compared to~\cite{Kawai:1985xq}.} 
\eqa\label{Kdef}
\nonumber &&K^{ss}\sim\frac{s^2}{4}(\epsilon\cdot\epsilon)^2-s(\epsilon\cdot k)^2(\epsilon\cdot\epsilon),\\
&&K\sim s(\epsilon\cdot k)^4-\frac{s^2}{4}\frac{1}{s+1}[s(\epsilon\cdot\epsilon)^2+(\epsilon\cdot k)^2(\epsilon\cdot\epsilon)-(\epsilon\cdot k)^4],
\eqae
where we've denoted only the degree of each type of Lorentz invariant. For the superstring one retains only $K^{ss}$, while the factor $K$ contains the tachyon poles for the bosonic string. The local gauge invariant function $K^{ss}$ corresponds to the $F^4$ operator in the effective action of open superstring. The pure graviton amplitude is given by 
\eq
\frac{(K^{ss})^2}{stu}\,.
\eqe

With the chiral boundary condition, one finds
\eq\label{Chiral}
M_{\rm 4, bos}^{\rm chiral}=-\frac{\pi}{stu}(\overline{K}^{ss}+\overline{K})(K^{ss}+K)
\eqe 
where $\overline{K}$ and $\overline{K}^{ss}$ indicates that all Lorentz invariants obtain a negative sign. The canceling of massive poles is straightforward to see at four-points. For the superstring, the open string on each side of the KLT formula has two sets of poles and a set of zeros. The flipping of signs then exchanges the physical and unphysical nature of the poles and zeros, thus resulting in mutual cancelation. This is illustrated diagrammatically below:\footnote{Note that while these diagrams are meant to show how the poles cancel, the space of kinematics is only two-dimensional due to the constraint $s+t+u=0$.}
$$\includegraphics[scale=0.5]{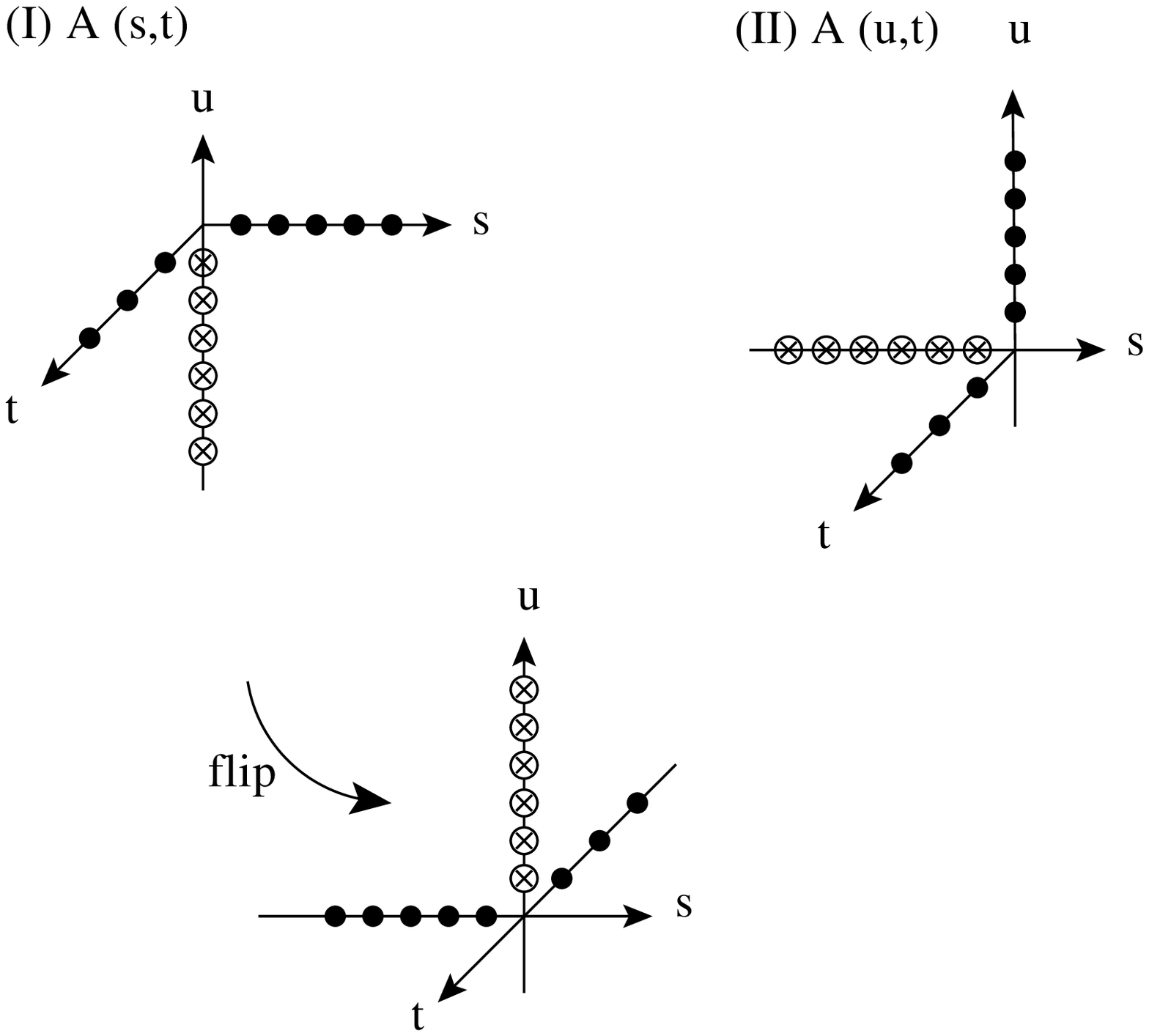}$$  
As shown above, the $u$ and $s$ channel poles cancel after the flip, while the $t$-channel now possesses poles both in the physical and unphysical channel. The latter is then canceled by the $\sin(\pi t)$ factor in the KLT kernel. When one of the factors is the bosonic string, the tachyon poles and the first massive pole simply exchange role upon the flip, and do not lead to cancelation. Thus the full amplitude contains only poles at $m^2=\pm1,0$. This is to be compared with the usual bosonic string four-point amplitude:
\eq
M_{\rm 4,bos}=\pi\frac{\G(-s)\G(-u)\G(-t)}{\G(1+s)\G(1+u)\G(1+t)} (K^{ss}+K)(K^{ss}+K)\,.
\eqe
Note that $K$ are of higher mass dimension compared to $K^{ss}$, while the latter are of the same degree in Lorentz invariants. Thus the change in sign does not effect the leading term in the $\alpha'$ expansion. In other words one recovers the pure graviton four-point amplitude. The same applies to the heterotic string. 

The residues of the massless poles in \eqref{Chiral} are the same as that of the bosonic string, with extra signs for terms of odd degree in Lorentz invariants. By dimension counting, one can see that the extra signs are consistent with the extra sign of $R^2$ operator that contributes to the three-point amplitude. For the residue of the massive pole, we consider the heterotic and bosonic string separately. 
\subsubsection{Heterotic \texorpdfstring{$\overline{\bf susy}\times$}{susy x} bosonic}
We now consider the massive poles in the heterotic string. For the case where the flipping of sign occurs in the supersymmetric sector, the amplitude is given by 
\eq
M_{\rm het}^{\rm chiral}=-\frac{\pi}{stu}(\overline{K}^{ss})(K^{ss}+K)\,.
\eqe
Due to the tachyon pole in $K$, the spectrum contains a tachyon whose residue is given by 
\eq\label{HetRes}
\frac{\pi}{t(-t+1)}(\overline{K}^{ss})\hat{K}|_{s=-1}\,,
\eqe
where $\hat{K}$ is the numerator of the tachyon pole in $K$ in \eqref{Kdef}. One can compare this to the residues of an open string tachyon and spin-2 particle:
\eqa\label{StringRes}
\nonumber \underset{s=-1}{\text{Res}}\left[\frac{\G[-s]\G[-t]}{\G[1+u]}(K^{ss}+K)\right]&=&-\frac{1}{t(-t+1)}\hat{K}|_{s=-1}\\
\underset{s=1}{\text{Res}}\left[\frac{\G[-s]\G[-t]}{\G[1+u]}(K^{ss})\right]&=&(K^{ss})|_{s=1}
\eqae
Looking back at \eqref{HetRes}, we see that the residue of our tachyon is \textit{minus} that of the product of open string tachyon and spin-2 residues, where the latter is evaluated with an extra sign for all inner products. This shows that the tachyon is indeed a spin-2 ghost field.  
\subsubsection{Heterotic \texorpdfstring{$\overline{\bf bosonic}\times$}{bosonic x} susy}
For the case where it is the bosonic sector that is flipped, we have a massive particle whose residue is given by 
\eq\label{Target}
-\frac{\pi}{t(t+1)}K^{ss}\overline{\hat{K}}|_{s=1}
\eqe
Comparing this to \eqref{StringRes}, we see the residue for the massive pole is simply the product of the open superstring spin-2 residue and the open bosonic string tachyon residue, where the latter is evaluated on a flipped signature. To see this, note that under flipped kinematics, the part of the bosonic string amplitude that contains the tachyon pole becomes: 
\eq
\left[\frac{\G[-s]\G[-t]}{\G[1+u]}\frac{\hat{K}}{s+1}\right]_{\rm filp}=\frac{\G[s]\G[t]}{\G[1-u]} \frac{\overline{\hat{K}}}{-s+1}\,,
\eqe
whose residue for the massive pole is $-\frac{\overline{\hat{K}}}{t(t+1)}$. This multiplied with the residue of the spin-2 state in the open superstring amplitude, $K^{ss}$, reproduces eq.(\ref{Target}). Since the relative sign is positive, this is a physical spin-2 state. 

\subsubsection{\texorpdfstring{$\overline{\bf bosonic}\times$}{bosonic x} bosonic}
For the bosonic case, both massive particles exists, and their residues are given as:
\eqa
\nonumber \underset{s=-1}{\text{Res}}\left[M_{\rm 4,bos}^{\rm chiral}\right]&=&\frac{\pi}{t(-t+1)}K (\overline{K}^{ss}+\overline{K})|_{s=-1}\\
\underset{s=1}{\text{Res}}\left[M_{\rm 4,bos}^{\rm chiral}\right]&=&-\frac{\pi}{t(t+1)}\overline{\hat{K}}(K^{ss}+K)|_{s=1}
\eqae
Since the leading $\alpha'$ expansion of the residue is identical with the of heterotic string, the nature of the massive states (physical or unphysical) stays the same.

\subsection{Massive external particles}

As shown in the previous subsection, apart from the massless states there is also a tachyonic state and a tardyonic state in the spectrum of the bosonic closed string, both of which are spin-2 particles. These states are forbidden in the ordinary bosonic string but are allowed by the chiral boundary condition, and an interesting question is whether the worldsheet SL(2) invariance is preserved when massive states are present in external lines. As a simple check, we look at the three-point amplitude of one tachyon, one graviton, and one tardyon. Let them be particles $1$, $2$ and $3$, respectively. Their vertex operators read
\begin{equation}
V_1=:\epsilon_1^{\mu\nu}\bar{\partial} X_\mu\bar{\partial} X_\nu e^{ik\cdot X}:,\qquad
V_2=:\epsilon_2^{\mu\nu}\partial X_\mu\bar{\partial}X_\nu e^{ik\cdot X}:,\qquad
V_3=:\epsilon_3^{\mu\nu}\partial X_\mu\partial X_\nu e^{ik\cdot X}:,
\end{equation}
where we flip the right movers. Note that when massive external states are involved the Koba--Nielsen factor carries non-trivial conformal weight, and the chiral boundary condition seems to change this weight for the right movers as compared to that of the ordinary string, thus making the three-point amplitude inconsistent. However, the state appearing on a given mass level is also changed (e.g., the tachyon switches from a scalar to a spin two), which guarantees that the amplitude is again invariant under the M\"obius transformation. This has been checked explicitly and the resulting amplitude is given by the product of the following tachyon-vector-spin-2 3-point open string amplitudes: 
\begin{align}
\nonumber A_{3L}&=k_1\cdot\epsilon_2\,k_{1,\mu}k_{1,\nu}\epsilon_3^{\mu\nu}-(k_{1,\mu}\epsilon_{2,\nu}+k_{1,\nu}\epsilon_{2,\mu})\epsilon_3^{\mu\nu},\\
A_{3R}&=-k_3\cdot\epsilon_2\,k_{3,\mu}k_{3,\nu}\epsilon_1^{\mu\nu}-(k_{3,\mu}\epsilon_{2,\nu}+k_{3,\nu}\epsilon_{2,\mu})\epsilon_1^{\mu\nu}.
\end{align}

Another check is whether with external massive states, the amplitude still contains only $m^2=\pm1,0$ poles. To see this it is useful to compare the poles and zeros of the open string between massive and massless external states. For massive ones, the difference is that while certain massless and tachyon poles might be absent, the zeros will extend from unphysical to physical region. Thus upon flipping the sign, the resulting amplitude can only have further pole cancelation, not less. Thus one concludes that only $m^2=\pm1,0$ poles can be present even for massive external states.

\section{Conclusion and discussions}

\subsection{Amplitudes}

In this paper we examined amplitudes in string theories modified by a chiral boundary condition. The new boundary condition leads to a simple operation acting on one of the open string sectors in the KLT representation, by flipping the sign of the spacetime metric. As a consequence, the closed string amplitudes show a remarkable cancelation among the infinite tower of massive poles, leaving only the massless poles in the type II superstring case, and poles in the adjacent massive levels in the bosonic and heterotic cases. These are consistent with the truncation of the spectrum caused by the chiral boundary condition. In particular, in the case of type II where only massless states are present, the interaction exactly reduces to that of type II supergravity, and we provide a general argument for it based on previous observations about the structure of open string amplitudes.

While in this paper we demonstrated that the amplitudes of the chiral string produce the desired field theory amplitudes, the argument relies on a non-trivial cancelation among an infinite number of massive poles in the KLT formula. It would be interesting to seek for a representation for these amplitudes manifesting the fact that the result is just a rational function of the kinematics data. A possible solution to this is to see how this evaluation of the amplitudes relates to the scattering-equation expressions that motivated it.  Perhaps a different choice of integration contour can be found that relates to the singular gauge previously used, and to the intermediate gauges that connect it to the conformal gauge.

\subsection{Actions}

There are several reasons why tachyons must have ghosts in Hermitian actions when their spin is nonvanishing.  This is clear in the case of supersymmetry from the usual positive-energy arguments; but it also follows more generally for fermions, since they satisfy first-order differential equations with real ``mass", so the relative signs for the two terms in the Klein--Gordon equation are always $\Box-\kappa$ for some nonnegative number $\kappa$ after squaring.  For positive-spin bosons, it follows from the use of St\"uckelberg fields, since changing the sign of the mass term also changes that for the two-derivative term of the St\"uckelberg fields; this means that either the longitudinal or transverse modes are ghosts.  These problems are related to the fact that generating tachyonic mass terms by dimensional reduction requires reducing a timelike direction.

However, our sign-flipping prescription can imply non-Hermitian actions:  For example, changing the sign of the metric implies the corresponding Dirac matrices get an extra factor of ``$i$", since $\{\gamma,\gamma\}$ changes sign.  Also, the normalization of external vectors in that sector gets an extra sign, suggesting a field redefinition with a factor of $i$.

Since the field theory action of \cite{Hohm:2013jaa} already contains the massive fields, it may well be that its Feynman diagrams produce amplitudes in exact agreement with those of the modification of (bosonic) string theory considered here.  Its potential shows how these masses are generated by spontaneous breaking of T-duality symmetry, with the massless fields as Goldstone bosons.  This breaking is the same as such breaking in polynomial $\sigma$-models, rather than in coset models, in that the potential is cubic and breaking also generates massive fields.  (There is also a Higgs mechanism involved, but in this case the non-Goldstones do the eating of vectors, rather than vice versa.)

But the two-derivative term there (as easily analyzed in a Landau gauge, or just ignoring divergence terms) comes with the wrong sign for {\it both} massive fields.  This again suggests a field redefinition with an $i$ to compare our amplitudes with the real action given there.

It would also be interesting to generalize that action to the heterotic case:  In particular, in the case without tachyons, it would probably provide a consistent description of physical, massive supergravity coupled to massless supergravity.

\section*{Acknowledgements}

\vspace{-0.3cm}

We are grateful to Oliver Schlotterer for enlightening discussions. Y-t.\,H.~is supported by MOST under the grant No.~103-2112-M-002-025-MY3, and would like to thank the Institute for Advanced Study for kind hospitality during the completion of this work. 
W.\,S.~is supported in part by National Science Foundation Grant No.\ PHY-1316617. E.\,Y.\,Y.~is supported by the U.S.~Department of Energy under grant DE-SC0009988, and by a Corning Glass Works Foundation Fellowship Fund at the Institute for Advanced Study.

\vskip .3 cm

\bibliographystyle{utphys}
\bibliography{leftstring}

\providecommand{\href}[2]{#2}\begingroup\raggedright\begin{thebibliography}{10}

\bibitem{Cachazo:2013hca}
F.~Cachazo, S.~He, and E.~Y. Yuan, ``{Scattering of Massless Particles in
  Arbitrary Dimensions},''
  \href{http://dx.doi.org/10.1103/PhysRevLett.113.171601}{{\em Phys. Rev.
  Lett.} {\bfseries 113} no.~17, (2014) 171601},
\href{http://arxiv.org/abs/1307.2199}{{\ttfamily arXiv:1307.2199 [hep-th]}}.

\bibitem{Cachazo:2013iea}
F.~Cachazo, S.~He, and E.~Y. Yuan, ``{Scattering of Massless Particles:
  Scalars, Gluons and Gravitons},''
  \href{http://dx.doi.org/10.1007/JHEP07(2014)033}{{\em JHEP} {\bfseries 07}
  (2014) 033},
\href{http://arxiv.org/abs/1309.0885}{{\ttfamily arXiv:1309.0885 [hep-th]}}.

\bibitem{Cachazo:2014nsa}
F.~Cachazo, S.~He, and E.~Y. Yuan, ``{Einstein-Yang-Mills Scattering Amplitudes
  From Scattering Equations},''
  \href{http://dx.doi.org/10.1007/JHEP01(2015)121}{{\em JHEP} {\bfseries 01}
  (2015) 121},
\href{http://arxiv.org/abs/1409.8256}{{\ttfamily arXiv:1409.8256 [hep-th]}}.

\bibitem{Cachazo:2014xea}
F.~Cachazo, S.~He, and E.~Y. Yuan, ``{Scattering Equations and Matrices: From
  Einstein To Yang-Mills, DBI and NLSM},''
  \href{http://dx.doi.org/10.1007/JHEP07(2015)149}{{\em JHEP} {\bfseries 07}
  (2015) 149},
\href{http://arxiv.org/abs/1412.3479}{{\ttfamily arXiv:1412.3479 [hep-th]}}.

\bibitem{Gross:1987kza}
D.~J. Gross and P.~F. Mende, ``{The High-Energy Behavior of String Scattering
  Amplitudes},''
\href{http://dx.doi.org/10.1016/0370-2693(87)90355-8}{{\em Phys. Lett.}
  {\bfseries B197} (1987) 129}.

\bibitem{Mason:2013sva}
L.~Mason and D.~Skinner, ``{Ambitwistor strings and the scattering
  equations},'' \href{http://dx.doi.org/10.1007/JHEP07(2014)048}{{\em JHEP}
  {\bfseries 07} (2014) 048},
\href{http://arxiv.org/abs/1311.2564}{{\ttfamily arXiv:1311.2564 [hep-th]}}.

\bibitem{Ohmori:2015sha}
K.~Ohmori, ``{Worldsheet Geometries of Ambitwistor String},''
  \href{http://dx.doi.org/10.1007/JHEP06(2015)075}{{\em JHEP} {\bfseries 06}
  (2015) 075},
\href{http://arxiv.org/abs/1504.02675}{{\ttfamily arXiv:1504.02675 [hep-th]}}.

\bibitem{Casali:2015vta}
E.~Casali, Y.~Geyer, L.~Mason, R.~Monteiro, and K.~A. Roehrig, ``{New
  Ambitwistor String Theories},''
  \href{http://dx.doi.org/10.1007/JHEP11(2015)038}{{\em JHEP} {\bfseries 11}
  (2015) 038},
\href{http://arxiv.org/abs/1506.08771}{{\ttfamily arXiv:1506.08771 [hep-th]}}.

\bibitem{Bjerrum-Bohr:2014qwa}
N.~E.~J. Bjerrum-Bohr, P.~H. Damgaard, P.~Tourkine, and P.~Vanhove,
  ``{Scattering Equations and String Theory Amplitudes},''
  \href{http://dx.doi.org/10.1103/PhysRevD.90.106002}{{\em Phys. Rev.}
  {\bfseries D90} no.~10, (2014) 106002},
\href{http://arxiv.org/abs/1403.4553}{{\ttfamily arXiv:1403.4553 [hep-th]}}.

\bibitem{Siegel:2015axg}
W.~Siegel, ``{Amplitudes for left-handed strings},''
\href{http://arxiv.org/abs/1512.02569}{{\ttfamily arXiv:1512.02569 [hep-th]}}.

\bibitem{Hohm:2013jaa}
O.~Hohm, W.~Siegel, and B.~Zwiebach, ``{Doubled $\alpha'$-geometry},''
  \href{http://dx.doi.org/10.1007/JHEP02(2014)065}{{\em JHEP} {\bfseries 02}
  (2014) 065},
\href{http://arxiv.org/abs/1306.2970}{{\ttfamily arXiv:1306.2970 [hep-th]}}.

\bibitem{Kawai:1985xq}
H.~Kawai, D.~C. Lewellen, and S.~H.~H. Tye, ``{A Relation Between Tree
  Amplitudes of Closed and Open Strings},''
\href{http://dx.doi.org/10.1016/0550-3213(86)90362-7}{{\em Nucl. Phys.}
  {\bfseries B269} (1986) 1}.

\bibitem{BjerrumBohr:2010hn}
N.~E.~J. Bjerrum-Bohr, P.~H. Damgaard, T.~Sondergaard, and P.~Vanhove, ``{The
  Momentum Kernel of Gauge and Gravity Theories},''
  \href{http://dx.doi.org/10.1007/JHEP01(2011)001}{{\em JHEP} {\bfseries 01}
  (2011) 001},
\href{http://arxiv.org/abs/1010.3933}{{\ttfamily arXiv:1010.3933 [hep-th]}}.

\bibitem{Mafra:2011nv}
C.~R. Mafra, O.~Schlotterer, and S.~Stieberger, ``{Complete N-Point Superstring
  Disk Amplitude I. Pure Spinor Computation},''
  \href{http://dx.doi.org/10.1016/j.nuclphysb.2013.04.023}{{\em Nucl. Phys.}
  {\bfseries B873} (2013) 419--460},
\href{http://arxiv.org/abs/1106.2645}{{\ttfamily arXiv:1106.2645 [hep-th]}}.

\bibitem{Mafra:2011nw}
C.~R. Mafra, O.~Schlotterer, and S.~Stieberger, ``{Complete N-Point Superstring
  Disk Amplitude II. Amplitude and Hypergeometric Function Structure},''
  \href{http://dx.doi.org/10.1016/j.nuclphysb.2013.04.022}{{\em Nucl. Phys.}
  {\bfseries B873} (2013) 461--513},
\href{http://arxiv.org/abs/1106.2646}{{\ttfamily arXiv:1106.2646 [hep-th]}}.

\bibitem{Schlotterer:2012ny}
O.~Schlotterer and S.~Stieberger, ``{Motivic Multiple Zeta Values and
  Superstring Amplitudes},''
  \href{http://dx.doi.org/10.1088/1751-8113/46/47/475401}{{\em J. Phys.}
  {\bfseries A46} (2013) 475401},
\href{http://arxiv.org/abs/1205.1516}{{\ttfamily arXiv:1205.1516 [hep-th]}}.

\bibitem{BroedelWebsite}
J.~Broedel, O.~Schlotterer, and S.~Stieberger, ``$\alpha'$ -expansion of open
  superstring amplitudes.''
\newblock \url{http://wwwth.mpp.mpg.de/members/stieberg/mzv/index.html}.

\bibitem{Broedel:2013tta}
J.~Broedel, O.~Schlotterer, and S.~Stieberger, ``{Polylogarithms, Multiple Zeta
  Values and Superstring Amplitudes},''
  \href{http://dx.doi.org/10.1002/prop.201300019}{{\em Fortsch. Phys.}
  {\bfseries 61} (2013) 812--870},
\href{http://arxiv.org/abs/1304.7267}{{\ttfamily arXiv:1304.7267 [hep-th]}}.

\bibitem{Broedel:2013aza}
J.~Broedel, O.~Schlotterer, S.~Stieberger, and T.~Terasoma, ``{All order
  $\alpha'$-expansion of superstring trees from the Drinfeld associator},''
  \href{http://dx.doi.org/10.1103/PhysRevD.89.066014}{{\em Phys. Rev.}
  {\bfseries D89} no.~6, (2014) 066014},
\href{http://arxiv.org/abs/1304.7304}{{\ttfamily arXiv:1304.7304 [hep-th]}}.

\bibitem{Brown:2011ik}
F.~Brown, ``{On the decomposition of motivic multiple zeta values},''
\href{http://arxiv.org/abs/1102.1310}{{\ttfamily arXiv:1102.1310 [math.NT]}}.

\bibitem{Huang:2016tag}
Y.-t. Huang, O.~Schlotterer, and C.~Wen, ``{Universality in string
  interactions},''
\href{http://arxiv.org/abs/1602.01674}{{\ttfamily arXiv:1602.01674 [hep-th]}}.

\bibitem{Naseer:2016izx}
U.~Naseer and B.~Zwiebach, ``{Three-point Functions in Duality-Invariant
  Higher-Derivative Gravity},''
\href{http://arxiv.org/abs/1602.01101}{{\ttfamily arXiv:1602.01101 [hep-th]}}.

\bibitem{Hohm:2014eba}
O.~Hohm and B.~Zwiebach, ``{Green-Schwarz mechanism and $\alpha'$-deformed
  Courant brackets},'' \href{http://dx.doi.org/10.1007/JHEP01(2015)012}{{\em
  JHEP} {\bfseries 01} (2015) 012},
\href{http://arxiv.org/abs/1407.0708}{{\ttfamily arXiv:1407.0708 [hep-th]}}.

\bibitem{Hohm:2015mka}
O.~Hohm and B.~Zwiebach, ``{Double Metric, Generalized Metric and
  $\alpha'$-Geometry},''
\href{http://arxiv.org/abs/1509.02930}{{\ttfamily arXiv:1509.02930 [hep-th]}}.

\end{thebibliography}\endgroup

\end{document}